# Implementing an Autonomous Emergency Braking with Simulink using two Radar Sensors.

Ritesh Kapse, Department of Electronics and Communication Engineering, Amrita School of Engineering,Coimbatore, Amrita Vishwa Vidyapeetham, Amrita University, India – 641112.

**Abstract:** In this paper we have implemented the autonomous emergency braking using two radar sensors with different angle of coverage. The synthetic radar data is generated by radar detection generator block available in AEBTestBench simulation module. AEBTestBench is autonomous emergency simulation module available in Matlab 2018b version under ADAS toolbox. From different EURO NCAP standard scenarios available, we have covered 5 scenarios with their results analysis showing where forward collision warning is displayed and what is the AEB status. The observation of simulation results are carried out for 10 seconds. Data fusion for two radar sensors is carried by Kalman filter algorithm present inside the simulation module.

**Keywords:** Inertial measurement unit (IMU), light-emitting diode detection and ranging (Leddar), Radio Detection And Ranging (RaDAR), Advanced driver-assistance systems (ADAS), European New Car Assessment Program (Euro NCAP), Autonomous emergency braking (AEB), Anti-Lock Braking System (ABS), FMCW (frequency-modulated continuous wave), MIO(Most Important Object), full braking (FB), partial braking (PB), time-to-collision (TTC)

## 1 Introduction

Autonomous emergency braking (AEB) is an ADAS active safety system that helps drivers avoid or decrease the impact of collisions with other vehicles or vulnerable road users. To automate dynamic driving tasks automated driving systems use combinations of sensor technologies like vision, radar, ultrasound etc. These tasks can involve steering control, braking control and acceleration control

AEB includes mainly below two points:

1. To warn the driver and identify critical scenarios can prevent accidents.
2. Some crashes are unavoidable, in such situations we can decrease the speed of vehicle in small steps before hard brakes applied to serve the minimal impact.

Most of the today's vehicles use radar and vision sensors for emergency braking to identify critical braking ahead of the ego vehicle. No sensor type works well for all the task in all conditions so sensor fusion will be necessary to provide redundancy for autonomous functions. Thus for the AEB system sensor fusion technology plays significant role AEB city and inter-urban system included the European New Car Assessment Program (Euro NCAP) for its safety rating from 2014. To protect most important road users such as pedestrians and cyclists, AEB systems are promoted continuously by EURO NCAP.We will be using Matlab Simulink 2018b version to implement automatic emergency braking (AEB) with a sensor fusion algorithm by using Automated Driving System Toolbox. In Matlab, we can test the AEB system in a closed-loop Simulink model using a series of test scenarios created by the Driving Scenario Designer app. In Driving Scenario Designer app, we can cover different standard scenarios given by EURO NCAP standards. Matlab Simulation can automatically generate C code for the control algorithm to implement the different functions.

For multisensor fusion development different algorithms such as Kalman filters, assignment algorithms, motion models, and a multiobject tracking framework etc. are supported in Automated Driving System Toolbox.

## 2 Sensor Details

### 1.1 RaDAR (Radio Detection and Ranging)

RaDAR sensors usually work on frequencies 24 or 76-81 GHz. The 24GHz systems are being used for short- and mid-range smart-driving features such as blind-spot detection and collision avoidance in a wider area. Ranges of those sensors are around 50 – 60 m with a beam angle of 90 degrees. The 24GHz frequency band has a few limitations. These include the potential for interference with radio astronomy and satellite services and, as a result, these RaDARs will be phased out of new vehicles by 2022 in Europe.

Unlike pulsed radar systems that are commonly seen in the defense industry, automotive radar systems often adopt FMCW (frequency-modulated continuous wave) technology. For smaller and cheaper radars with less power usage we can go for FMCW radars in comparison with pulsed radars. To monitor smaller distances, FMCW radars are best choice.

Range and Doppler estimation through computationally efficient FFT operations are enabled by FMCW waveforms. In most of the automotive radars, center frequency of operation is 77 GHz When radar is mounted on ego vehicle, for long range operations, it should detect the objects or vehicles till range of 100 meters. Objects in range that are at least 1 meter apart, are required to be resolved by radar. Since this radar is forward facing, the targets with closing speed as high as 230 km/hr should be handled by radar.

Following figure shows the principle of range measurement using the FMCW technique.

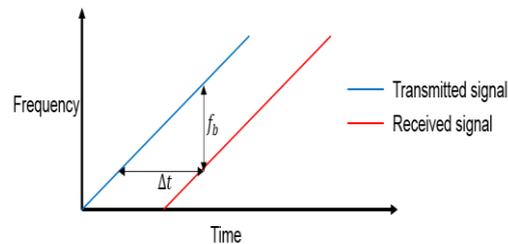
Fig1 : Range measurement using FMCW technique.

When transmitted signal is time-delayed, it is same as received signal where $\Delta t$ is related to range. Since signal is sweeping through frequency band always, frequency difference *fb* is constant between transmitted and received signal at any moment during the sweep. *fb* is nothing but the beat frequency. We can get the time delay from beat frequency since sweep is linear and then delay can be translated to the range.

Up-sweep linear FMCW signal, often referred to as saw tooth shape is shown in below figure.

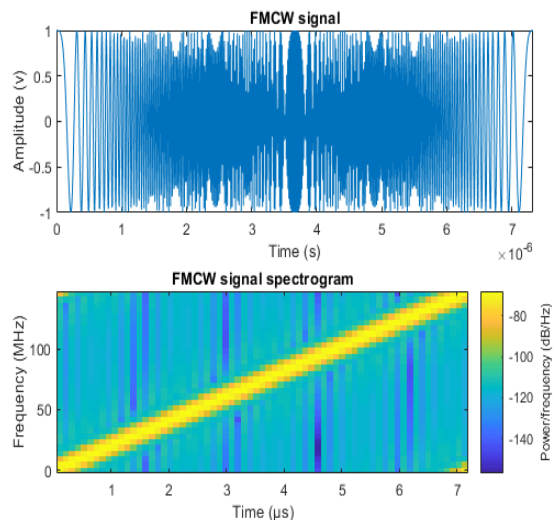
Fig 2: FMWC signal and its spectrogram

Beat frequency is examined by FMCW radar in the dechirped signal to get range measurements. Received signal is mixed with transmitted signal to perform dechirp operation so that beat frequency is extracted. Now individual frequency components which corresponds to target range are only available in the dechirped signal after mixing is completed.

From a single sweep it is possible to extract the Doppler information. But within one pulse, the Doppler frequency is indistinguishable from the beat frequency. So the Doppler shift is generally extracted within several sweeps.

Below operations are carried out by FMCW radar to measure the range and Doppler.

1. The FMCW signal is generated by waveform generator.
2. The signal is amplified by transmitter and the antenna and then radiated into space.
3. The signal broadcasts to the target, gets reflected by the target, and travels back to the radar.
4. The signal is collected by the receiving antenna.
5. Once signal is dechirped, the received signal is saved in a buffer.
6. After certain number of sweeps are filled in buffer, the beat frequency as well as the Doppler shift is extracted by performing Fourier transform on range and Doppler. Hence range and speed of target can be estimated using these results.

## 3 Simulation

For simulation using AEBTestBench module, we will be using two radar sensors with different angular view so that field of coverage will be more during simulation.
There are two main subsystems in model:
- First one is AEB with Sensor Fusion that covers the sensor fusion algorithm and AEB controller.
- Second is Vehicle and Environment, which replicates the ego vehicle dynamics and the environment. Synthetic sensor data for the objects are provided by driving scenario reader and radar detection generators.

Stopping time is calculated and accordingly forward collision warning (FCW) and AEB control algorithm are implemented by using AEB Controller. We will be running simulation for 10 seconds to check the all possible results.

From the time ego vehicle applies its brakes with deceleration $a_{brake}$ at first time, till it is completely stopped, that time period is called as stopping time.

Mathematically stopping time obtained by following equation:

$$\tau_{stop} = v_{ego}/a_{brake} \qquad (1)$$

When FCW system will alert drivers for imminent collision with a lead vehicle, it is expected to react to the alert and apply the brake with a delay time $\tau_{react}$.

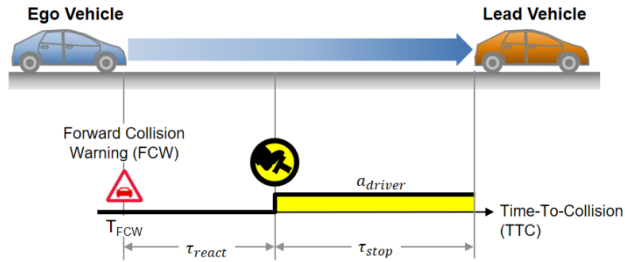

Fig 3: Delay time ($\tau_{react}$) and Stopping time ($\tau_{stop}$)

Total time taken by ego vehicle to travel before colliding with lead vehicle is given by:

$$T_{FCW} = \tau_{react} + \tau_{stop} = \tau_{react} + v_{ego}/a_{driver} \qquad (2)$$

The FCW alert is activated when the time-to-collision (TTC) of the lead vehicle is less than $T_{FCW}$

Sometimes due to distractions driver may failed to apply brakes, at such situations to avoid or mitigate the collision the AEB system acts independently. AEB system typically applies step by step braking where the multi-stage partial braking is followed by full braking.

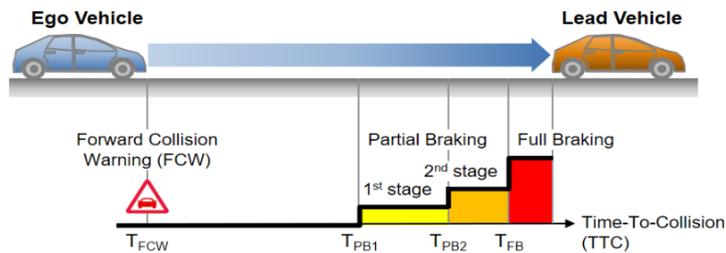

Fig 4: FCW and partial braking activation in critical scenarios.

Autonomous emergency braking logic used in AEB controller to activate the forward collision warning and display the AEB status is as follows:

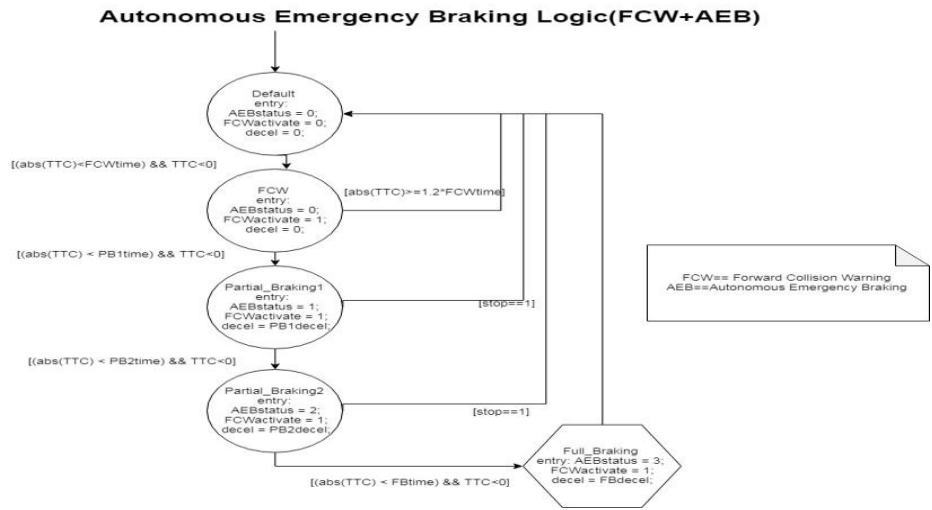

Fig 5: Autonomous Emergency Braking Logic

Different blocks in AEB controller with AEB logic has:
TTCCalculation calculates the TTC using the relative distance and velocity of the lead vehicle or the most important object
StoppingTimeCalculation calculates stopping times for FCW, first- and second-stage partial braking (PB), and full braking (FB)
FCW and AEB activations are determined by AEB_Logic by comparing the TTC with the stopping times.

## 4 Experimental Results

Total of five graphs with detailed analysis will be displayed in results according to different EURO NCAP standard scenarios.
- Comparison between time-to-collision (TTC) and the stopping times for the FCW, first stage partial brake, second stage partial brake and full brake is shown in the first plot (TTC vs. Stopping Time).
- Based on the comparison results from the first plot, how the AEB state machine determines the activations for FCW and AEB is shown in second plot.
- The velocity of the ego vehicle is displayed in third plot.
- The acceleration of the ego vehicle is shown in fourth plot.
- The headway between the ego vehicle and the MIO (Most Important Object) is displayed in fifth plot.

Five scenarios are simulation results are as follows:

- AEB_CCRs_50overlap

European standard scenario (EURO NCAP) that is a car-to-car rear stationary (CCRs) scenario, where the ego car rear-ends a stationary vehicle and 50% of the width of the moving vehicle is overlapped by ego vehicle at collision time.

After simulation, results for AEB and FCW are as follows

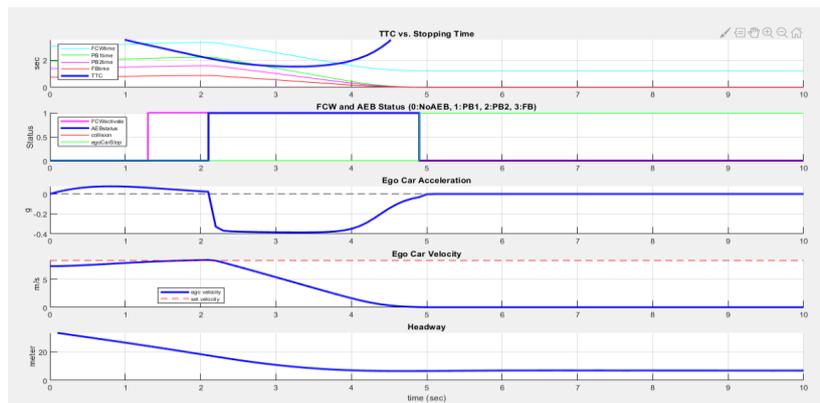

Fig 6 : AEB results for Standard EURO NCAP scenario 1

- AEB_CCRm_50overlap

European standard scenario (EURO NCAP) that is a car-to-car rear moving (CCRm) scenario, where the ego car rear-ends a moving vehicle and the ego car overlaps with 50% of the width of the moving vehicle at time of collision.

After simulation, results for AEB and FCW are as follows:

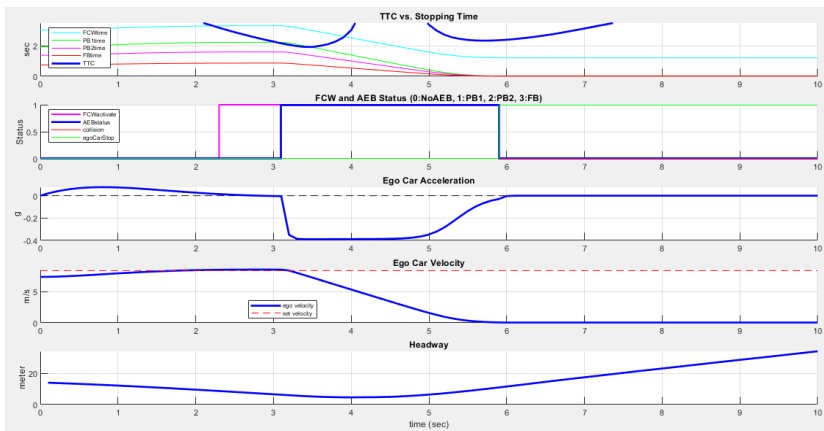

Fig 7 : AEB results for Standard EURO NCAP scenario 2

- AEB_CCRb_6_initialGap_12m

European standard scenario (EURO NCAP) that is a car-to-car rear braking (CCRb) scenario, where the ego car rear-ends a braking vehicle and the braking vehicle begins to decelerate at 6 m/s2. The ego car and the braking vehicle has initial gap of 12m

After simulation, results for AEB and FCW are as follows:

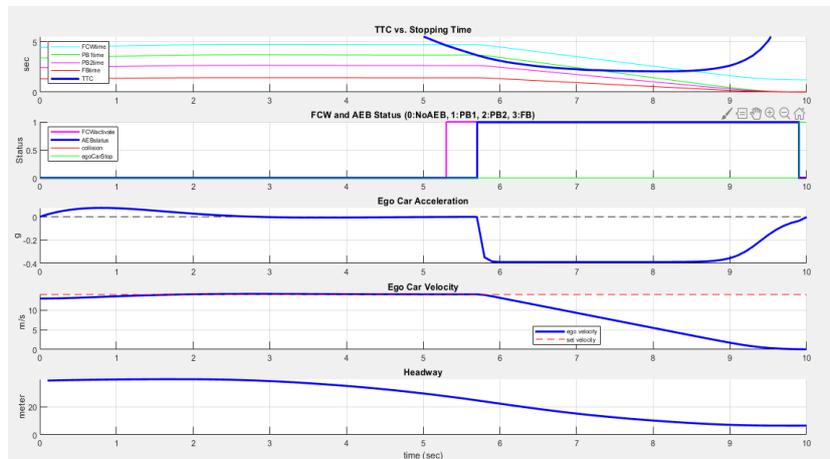

Fig 8 : AEB results for Standard EURO NCAP scenario 3

- AEB_Pedestrian_Nearside_25width

Euro NCAP test protocols refer to as the near side when the ego car collides with a pedestrian who is traveling from the right side of the road .According to these protocols it is assumed that vehicles travel on the right side of the road. Hence, the side nearest to the ego car is the right side of the road. The pedestrian is 25% of the way across the width of the ego car at collision time.

After simulation, results for AEB and FCW are as follows:

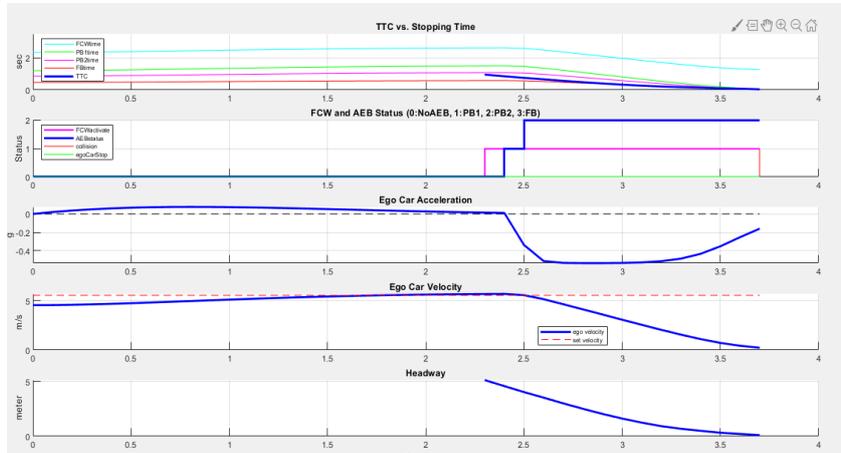

Fig 9 : AEB results for Standard EURO NCAP scenario 4

- AEB_CCRb_2_initialGap_40m

European standard scenario (EURO NCAP) that is a car-to-car rear braking (CCRb) scenario, where the ego car rear-ends a braking vehicle and the braking vehicle begins to decelerate at 2 m/s2. The ego car and the braking vehicle has initial gap of 40 m.

After simulation, results for AEB and FCW are as follows:

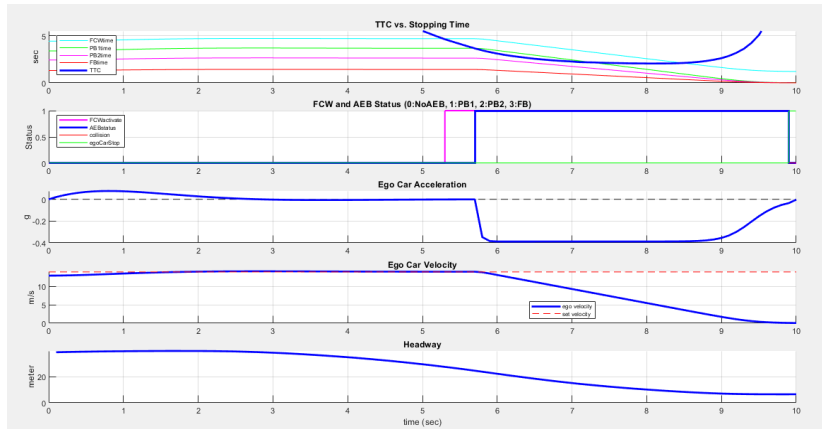

Fig 10: AEB results for Standard EURO NCAP scenario 5

## 5 Future Scope

The above experiment we will be performing with hardware available to crosscheck the synthetic data and actual data from radar sensor. There might be some errors due to efficiency on radar data in different environment conditions. So next step will be to minimize the errors and improve accuracy in comparison with actual and synthetic data. Currently the hardware we are available with are RaDAR ARW1243Boost and Leddar Tech (M16D-75B0005). We will implement the sensors on car and replicate the EURO NCAP scenarios to get data from both the sensors.

## 6 Conclusion

In this paper we have proposed, the autonomous emergency braking on ego vehicle using simulation module covering five standard scenarios defined by EURO NCAP. The result can clearly show the forward collision warning to driver, implementation of partial braking in steps and the automatic emergency braking.

All the above results are carried out using two RaDAR sensors synthetic data generated from RaDAR detection generator. Two different sets of data from RaDARs are combined by using Kalman Filter algorithm which is in-build to AEB Test Bench simulation module. In near future, it is necessary to have emergency braking module implemented in a vehicle. In European countries, it is already the most important safety measure to get five star rating to vehicles.